\documentclass[a4paper,final]{appolb}
\usepackage{epsfig}
\usepackage[font=rm, labelfont={rm,bf}, margin=1cm]{caption}

\begin{document}
\pagestyle{plain}
\title{Leptonic decay of $\phi$(1020) meson measured with the STAR experiment%
%
}
\author{Masayuki Wada for the STAR Collaboration
\address{Department of Physics, University of Texas at Austin}
}
\maketitle

\begin{abstract}
 We present the measurement of the $\phi$(1020)$\rightarrow e^{+}e^{-}$ transverse momentum spectrum and yield for $0.3 < p_{T} < 2.5$ GeV at mid-rapidity ($|y| < 1$) from the STAR experiment in Au+Au collisions at $\sqrt{s_{_{NN}}} = 200$ GeV and compare it to the yield and spectrum from the hadronic decays. We also compare the $\phi$(1020) mass and width obtained by fitting Breit-Wigner function to simulated values. Particle identification based on the newly installed STAR Time-Of-Flight (TOF) detector \cite{LLOPE2005} in conjunction with energy-loss (dE/dx) based particle identification measurements from the Time Projection Chamber (TPC) \cite{Ackermann2003} are used for a clean electron and positron identification. 
\end{abstract}

\section{Introduction}
Hadronic resonances can play a pivotal role in providing experimental evidence for partial chiral symmetry restoration \cite{Rapp1999a} in the deconfined quark-gluon phase produced at RHIC \cite{Adams2005} and the LHC. Their lifetimes, which are comparable to the lifetime of the fireball, make them a valuable tool to study medium modifications to the resonant state due to the chiral phase transition signatures of mass shifts and/or width broadenings. Because of the relatively small interaction of leptons with the hadronic medium, the resonances such as $\rho$(770), $\omega$(782), and $\phi$(1020) can be reconstructed via the leptonic decay channels without the hadronic medium effects at the later stage of the fireball expansion. However, hadronic regenerated resonances will decay leptonically as well. In this study, we focus on the $\phi$(1020) since its larger mass makes it well separated from the $\rho$ and the $\omega$ in the invariant mass space. 

\section{Analysis}
The data presented here was taken in the year 2010 from Au+Au collisions at $\sqrt{s_{_{NN}}} = 200$ GeV with the minimum bias trigger in the STAR experiment at RHIC.

Charged tracks are reconstructed using the TPC, which provides momentum, track path length, and the ionization energy loss information within its 2$\pi$ azimuthal coverage.  With the path length information, the newly installed TOF detector provides particle velocity information used for the particle identification. 

Events are selected with the position of the primary vertices along the beam axis (Vz) within $\pm$ 30 cm from the geometric center of the TPC. Since the TOF detector was used, we also required that the difference between the Vz reconstructed with the TPC and the Vz from the Vertex Position Detector (VPD), two TOF start side detectors, is within $\pm$ 13 cm.  
After an additional 0-80\% centrality selection based on multiplicity of charged particles at mid rapidity, about 250M events are analyzed in this study. 
 
Primary tracks which have a distance of closest approach (DCA) within 1.5 cm and have at least 15 fit points (NFitPts) in the TPC are selected. Tracks are also required to have the ratio of  the NFitPts to the number of possible point in the TPC (NPossPts), which is at most 45 points, be greater than 0.52. With this requirement we can avoid split tracks, one track reconstructed into two tracks.
 
\begin{figure}[h!tb]
\begin{center}
  \includegraphics[width=2.in]{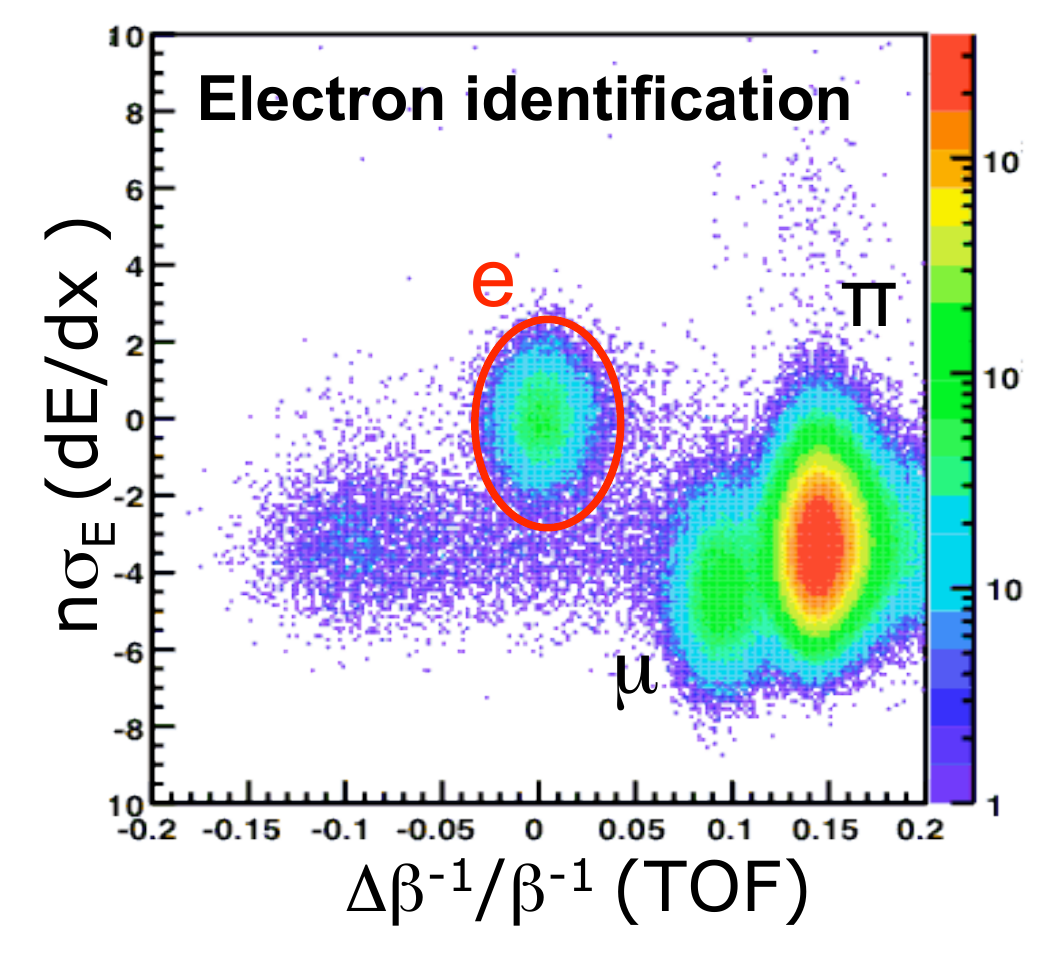}
  \caption[labelInTOC]{Combined electron identification (p=0.23-0.24 [GeV/c])}
    \label{fig:EID}
\end{center}
\end{figure}

As TOF and dE/dx measurements can identify particles independently, using them in combination allows for more separation power.
In order to obtain the probability of being electrons, we fill n$\sigma$(dE/dx), deviation from the mean in terms of standard deviation, and $\Delta\beta^{-1}/\beta^{-1} = \left(\beta^{-1}_{TOF}-\beta^{-1}_{expected}\right)/\beta^{-1}_{TOF}$ with the mass assumption of pion or electron as in the Fig. \ref{fig:EID}. These histograms are fit with a 2 dimensional Gaussian distributions for electrons and pions. Knowing the distributions, we can calculate the probability. In this analysis a probability bigger than 95\% is required to identify as an electron.
This method of applying dE/dx and $\beta$ cuts at the same time has an advantage over other methods in which the two cuts are applied separately because in the later case one cuts out a rectangle in the 2D histogram, by contrast, our method cuts out an ellipse with higher purity.   

The $\phi$ invariant mass is reconstructed from all $e^{+}$ and $e^{-}$ pairs in the same event (unlike-sign signal). Since these pairs include uncorrelated pairs as well, the mixed-event technique is adopted to estimate the combinatorial background. 

The mixed-event background is reconstructed with electron and positron pairs from different events. The mixed events are selected only from the same event class. We defined event classes with 10 bins in both Vz position and reaction plane angle of the events. After normalizing the mixed-events background to the unlike-sign signal and subtracting it from the unlike-sign signal, we can extract $\phi$ signals as in Fig. \ref{fig:InvMass} (right panel). To measure the $\phi$ transverse momentum spectrum we repeat the same procedure in each transverse momentum bin.

The spectrum is corrected for track reconstruction efficiency, TOF-hit matching efficiency, detector acceptance, and track quality cut efficiency as well as PID cut efficiency. The track reconstruction efficiency and TPC acceptance have been obtained by embedding Monte-Carlo (MC) tracks into real data at the detectors' response level and reconstructing them along with real tracks. We also apply the track quality cut to the reconstructed tracks to estimate the track quality cut efficiency. The TOF matching and acceptance efficiency are calculated from real data by comparing all reconstructed tracks with those that have associated TOF hits. The PID cut efficiency is calculated directory from the fit 2 dimensional Gaussian distributions mentioned above. 

The systematic uncertainties of the $\phi \rightarrow e^{+}e^{-}$ yield were estimated by varying the track selection cuts such as DCA, NFitPts, the ratio of the Nfitpts to the NPossPts, and the PID probability cut value. We also varied the normalization ranges for the mixed-events, the signal fit range, fit function form for the residual background, and histogram binning. The systematic errors are 15$\sim$24\%. The largest error is about 24\% at the highest transverse momentum bin. The change in the ratio of the Nfitpts to the NPossPts and the PID probability cut value contributes most to the systematic errors.

\section{Results}
\begin{figure}[htb]
\begin{center}$
\begin{array}[B]{cc}
\includegraphics[width=2.in]{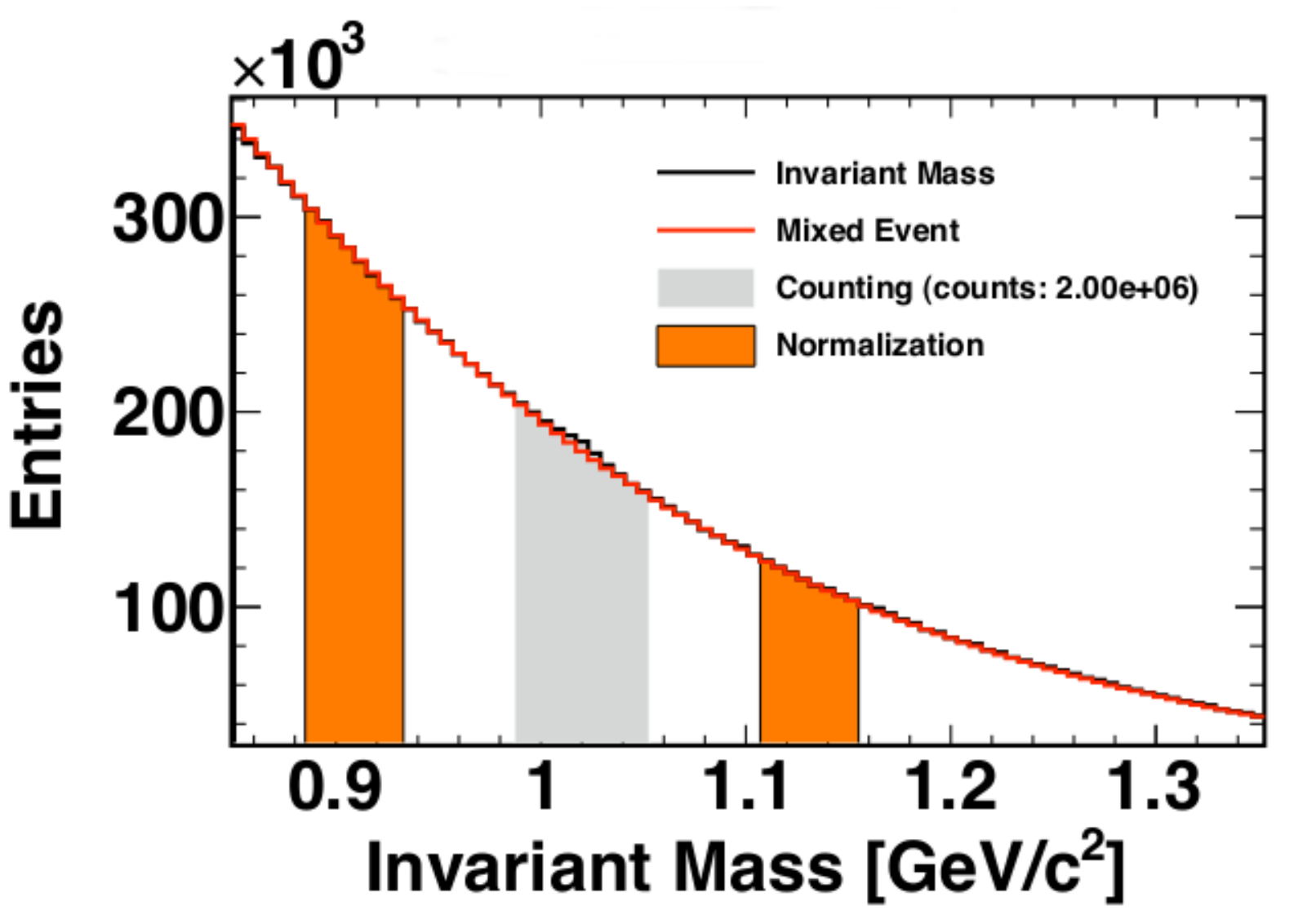} 
\includegraphics[width=2.1in]{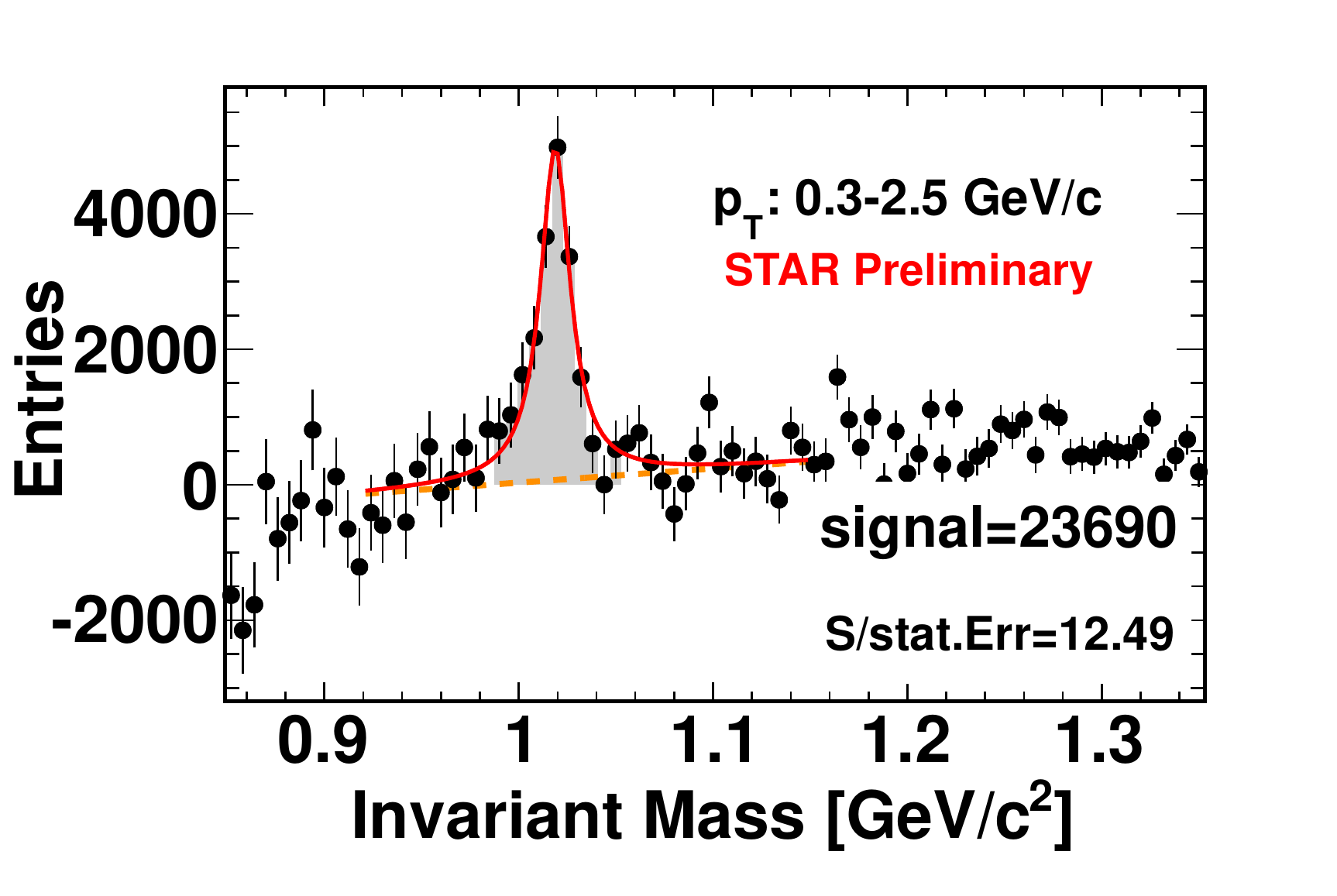} 
\end{array}$
\end{center}
\caption{Left: the invariant mass distribution of the $e^{+} e^{-}$ pairs in the same event (black line) with the normalized mixed-event background (red line) in transverse momentum range of 0.3 to 2.5 GeV/c. The orange areas show the normalization areas for the mixed-event background. 
Right: The $p_{T}$ integrated $\phi \rightarrow e^{+}e^{-}$ signals after the mixed-events background subtraction. The red lines are the non-relativistic Breit-Wigner functions representing the $\phi \rightarrow e^{+}e^{-}$ signals and the orange slopes show the residual backgrounds.}
\label{fig:InvMass}
\end{figure}

Figure \ref{fig:InvMass} (left panel) shows the invariant mass distribution of the $e^{+} e^{-}$ pairs in the same event (black line) with the normalized mixed-event background (red line) in the transverse momentum range of 0.3 to 2.5 GeV/c. The orange areas show the normalization areas for the mixed-event background. The integrated $\phi \rightarrow e^{+}e^{-}$ signals in the same $p_{T}$ range is presented in Fig. \ref{fig:InvMass} (right panel).  The red line is the non-relativistic Breit-Wigner function representing the $\phi \rightarrow e^{+}e^{-}$ signals and the orange line shows the residual background. The gray regions show the signal counting regions in both figures. We can extract the signal with significance of 12.5. 

%

\begin{figure}[h!tb]
\begin{center}$
\begin{array}{cc}
\includegraphics[width=2.in]{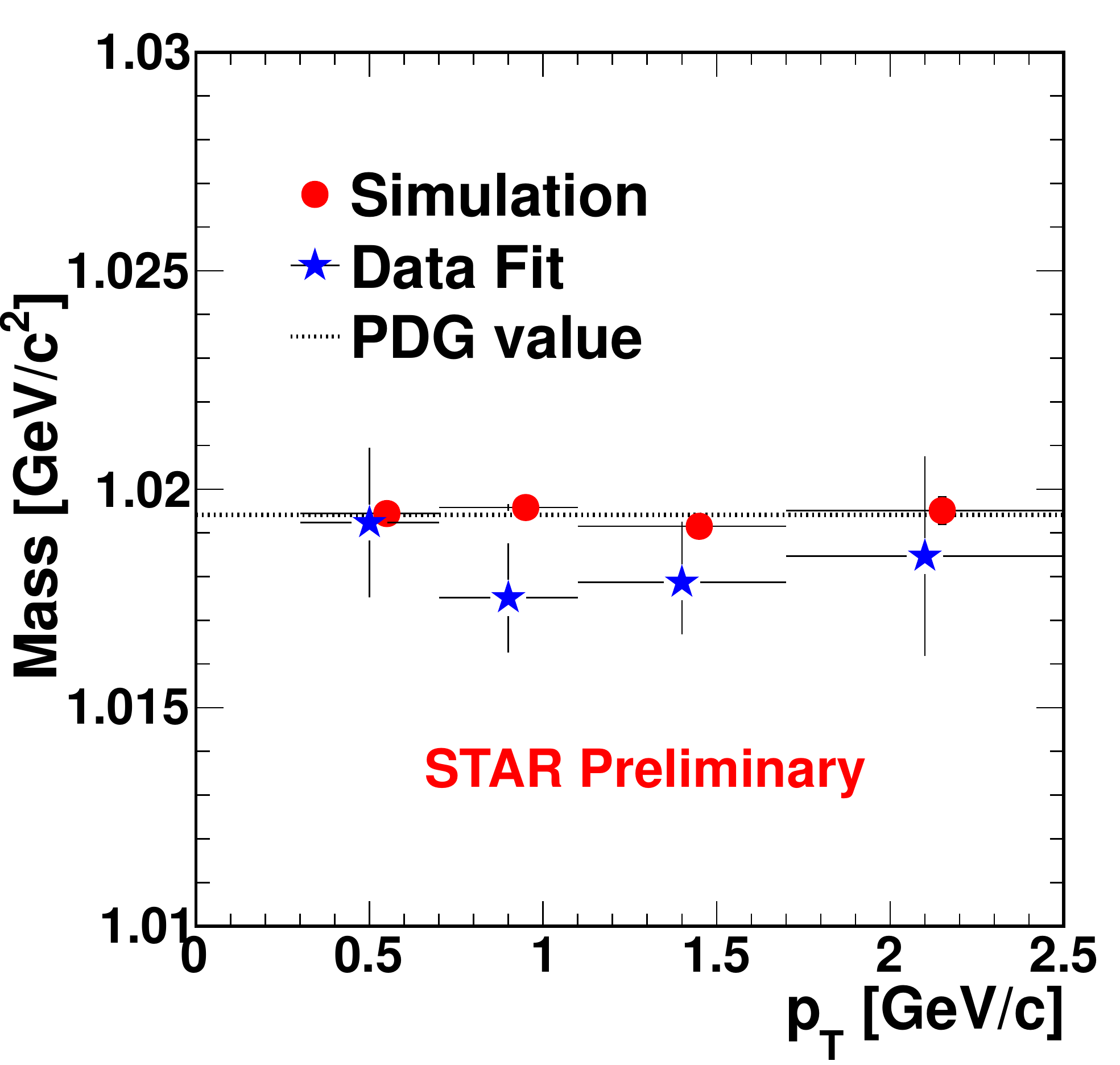} &
\includegraphics[width=2.in]{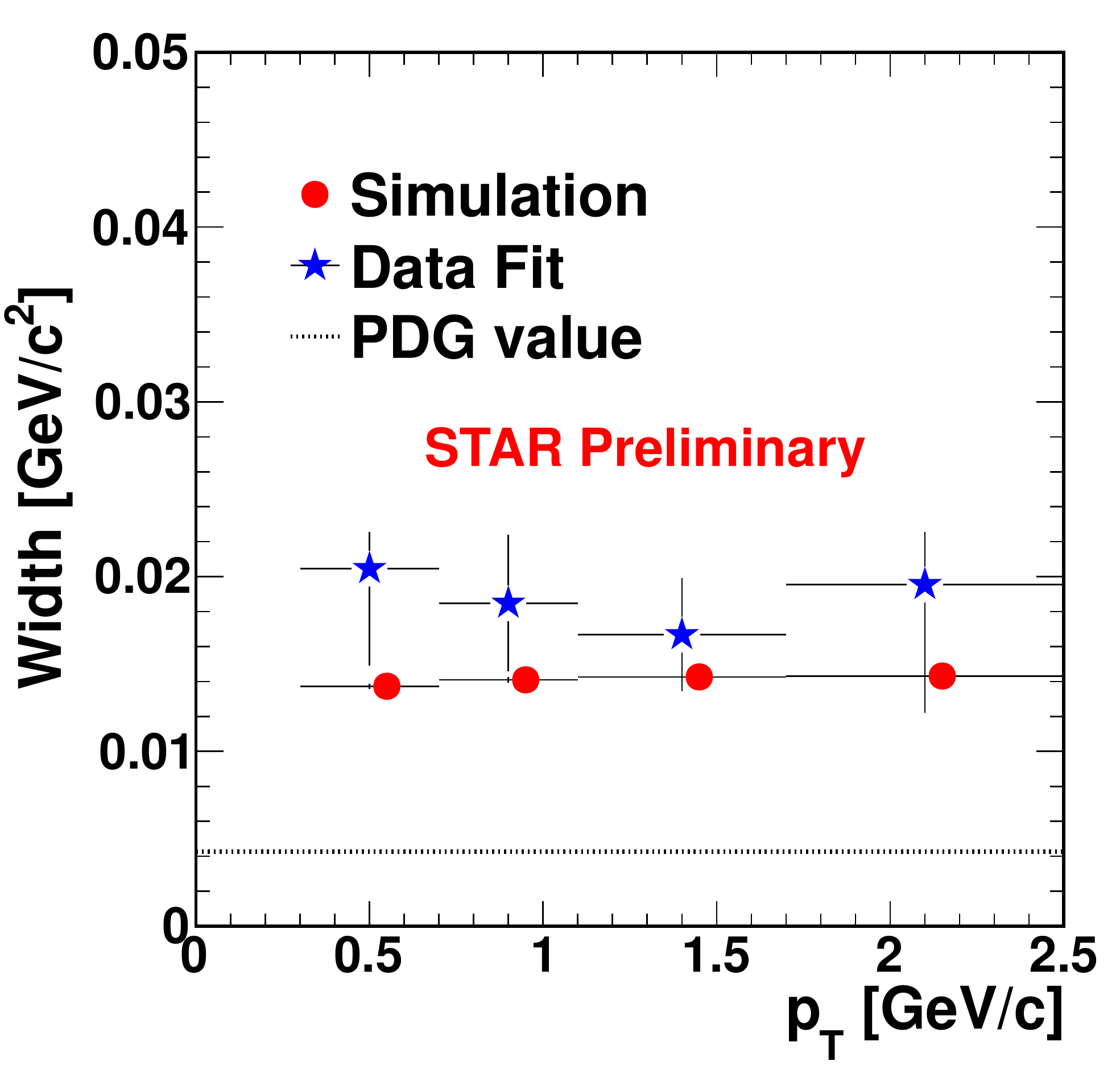} 
\end{array}$
\end{center}
\caption{Left: mass vs. transverse momentum of $\phi$(1020). 
Right: width vs. transverse momentum of $\phi$(1020). The blue points represent fitting values and the errors are only statistical. The  simulation results are plotted in red. The dashed lines show the PDG values.
}
\label{fig:MassWidth}
\end{figure}

Figure \ref{fig:MassWidth} shows the fit result of the mass and decay width of the $\phi$(1020) meson in each $p_{T}$ bin with blue points. The error bars are statistical errors (fit parameter errors) only. Simulation results and the particle data group (PDG) values \cite{PDG2010} are plotted with red points and dashed line respectively. This simulation results are obtained from the embedded MC tracks by fitting the invariant mass distribution after reconstruction. This means the results include the resolution of momentum measurement in STAR. The fit, simulation, and PDG values of the mass are consistent within $\sim1.5\sigma$. On the other hand, the fit decay widths show significant discrepancy from the PDG value. This width broadening, however, can be explained only by the momentum resolution of the detector because the simulation results also shows the broadening. 

\begin{figure}[h!tb]
\begin{center}
  \includegraphics[width=2.1in]{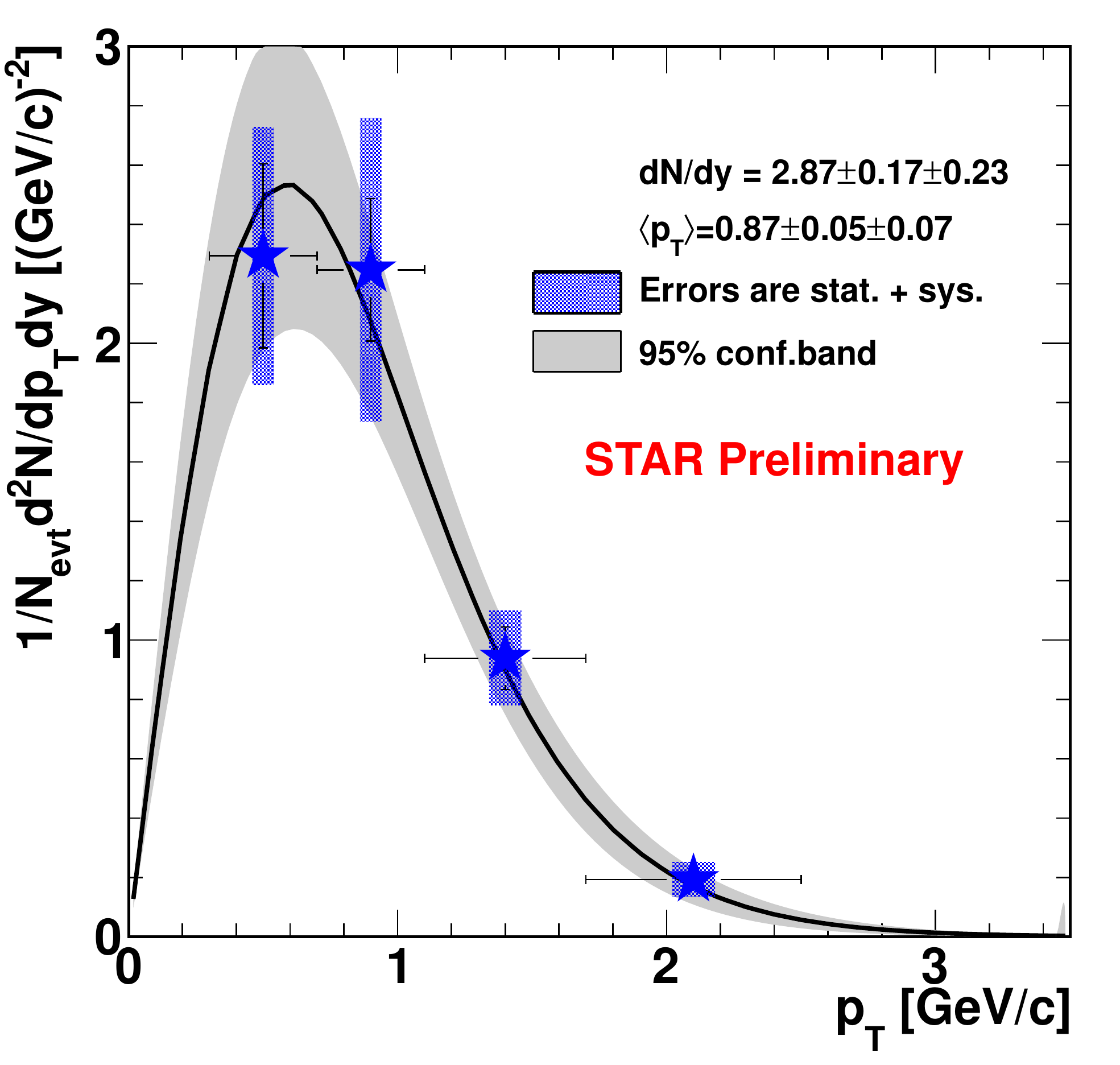}
  \caption[labelInTOC]{The corrected $\phi \rightarrow e^{+}e^{-}$ yields within rapidity $|y| < 1$ for the 0-80\% centrality events from Au + Au collisions at $\sqrt{s_{_{NN}}} = 200$ GeV. The vertical bars are statistical errors and the boxes are the sum of systematic and statistical errors in quadrature. The solid line is the exponential fit function and the gray band represents a 95\% confidence band of the fit. The values of $dN/dy$ and $\left< p_{T}\right>$ correspond to the best estimated values, statistic errors, and systematic errors respectively.}
    \label{fig:dNdydptFit}
  \includegraphics[width=2.1in]{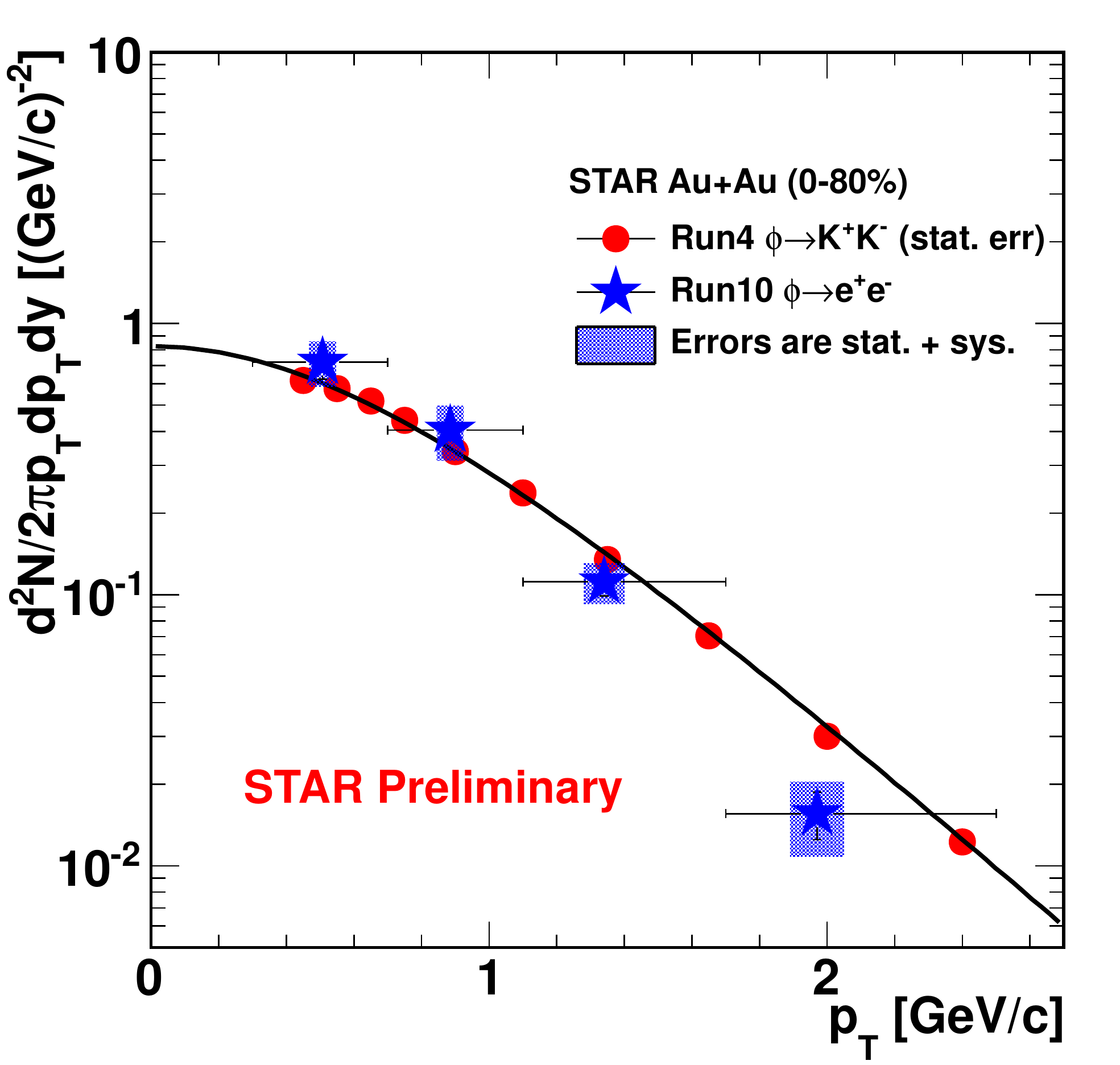}
  \caption[labelInTOC]{Comparison with the hadronic channel result \cite{Abelev2009}}   
   \label{fig:dNdydpt2pipt}
\end{center}
\end{figure}

The corrected $\phi \rightarrow e^{+}e^{-}$ yields at the rapidity $|y| < 1$ for the 0-80\% centrality events from Au + Au collisions at $\sqrt{s_{_{NN}}} = 200$ GeV are presented in Fig. \ref{fig:dNdydptFit}. The $\phi$(1020) yield per unit rapidity (dN/dy) is calculated by counting signals within the transverse momentum range and integrating an exponential fit function for the rest. The mean transverse momentum ($\left<p_{T}\right>$) is obtained in the similar way. The contributions from the extrapolation to unmeasured $p_{T}$ regions are about 13\% and 5\% for dN/dy and $\left<p_{T}\right>$ respectively. We obtain dN/dy=2.87$\pm$0.17(stat.)$\pm$0.23(sys.) and $\left< p_{T}\right>$ =0.87$\pm$0.05(stat.)$\pm$0.07(sys.) [GeV/c].


Figure \ref{fig:dNdydpt2pipt} shows $\phi \rightarrow e^{+}e^{-}$ invariant yields in blue and the red points represent STAR published $\phi \rightarrow K^{+}K^{-}$ results \cite{Abelev2009}. The error notations are the same as in the Fig. \ref{fig:dNdydptFit}. The $\phi$ measurements in di-leptonic decay channel are consistent with the STAR published results within 1.2$\sigma$ if we take into account the 10\% systematical uncertainty of the published result.   



\section{Summary}
The clear $\phi \rightarrow e^{+}e^{-}$ signal (significance $\sim$12.5) at rapidity $|y| < 1$ for the 0-80\% centrality events from Au + Au collisions at $\sqrt{s_{_{NN}}} = 200$ GeV is presented. The corrected $\phi$ transverse momentum spectrum is also shown. We obtain dN/dy = 2.87 $\pm$ 0.17(stat.) $\pm$ 0.23(sys.) and $\left< p_{T}\right>$ = 0.87 $\pm$ 0.05(stat.) $\pm$ 0.07(sys.) [GeV/c] from it. The $\phi \rightarrow e^{+}e^{-}$result is consistent with the previous $\phi \rightarrow K^{+}K^{-}$ result. No mass shift or width broadening beyond the known detector effects are observed. 

\bibliographystyle{phaip}
\bibliography{SQM}{}

\end{document}